\documentclass[aps,prl,preprint,tightenlines,superscriptaddress,,amsmath,byrevtex]{revtex4}
\usepackage{graphicx} 
\usepackage{dcolumn}  
\usepackage{float}
\usepackage{subfigure}
\usepackage{color,soul}
\usepackage{lineno}
\graphicspath{{ps}}

\RequirePackage{xspace}

\def\qqbar  {\ensuremath{q\overline q}\xspace}
\def\bbar   {\ensuremath{\overline b}\xspace}
\def\sbar   {\ensuremath{\overline s}\xspace}
\def\dbar   {\ensuremath{\overline d}\xspace}

\def\pip    {\ensuremath{\pi^+}\xspace}
\def\pim    {\ensuremath{\pi^-}\xspace}
\def\pipm   {\ensuremath{\pi^{\pm}}\xspace}
\def\Kbar   {\kern 0.2em\overline{\kern -0.2em K}{}\xspace}

\def\Kz     {\ensuremath{K^0}\xspace}
\def\Kzb    {\ensuremath{\Kbar^0}\xspace}
\def\KzKzb  {\ensuremath{\Kz \kern -0.16em \Kzb}\xspace}
\def\Kp     {\ensuremath{K^+}\xspace}
\def\Km     {\ensuremath{K^-}\xspace}

\def\kpm    {\ensuremath{K^{\pm}}\xspace}
\def\KS     {\ensuremath{K^0_{\scriptscriptstyle S}}\xspace}

\def\Dbar   {\kern 0.2em\overline{\kern -0.2em D}{}\xspace}

\def\Dz     {\ensuremath{D^0}\xspace}
\def\Dzb    {\ensuremath{\Dbar^0}\xspace}
\def\DzDzb  {\ensuremath{\Dz {\kern -0.16em \Dzb}}\xspace}
\def\Dp     {\ensuremath{D^+}\xspace}
\def\Dm     {\ensuremath{D^-}\xspace}
\def\DpDm   {\ensuremath{\Dp {\kern -0.16em \Dm}}\xspace}

\def\Dstarp {\ensuremath{D^{*+}}\xspace}

\def\Bbar   {\kern 0.18em\overline{\kern -0.18em B}{}\xspace}

\def\BB     {\ensuremath{B\Bbar}\xspace}

\def\Bu     {\ensuremath{B^+}\xspace}
\def\Bub    {\ensuremath{B^-}\xspace}
\def\Bp     {\ensuremath{\Bu}\xspace}
\def\Bm     {\ensuremath{\Bub}\xspace}
\def\Bpm    {\ensuremath{B^{\pm}}\xspace}
\def\hpm    {\ensuremath{h^{\pm}}\xspace}
\def\hu     {\ensuremath{h^+}\xspace}
\def\hp     {\ensuremath{\hu}\xspace}
\newcommand{\mkk}{\ensuremath{M_{K_{S}^{0}K_{S}^{0}}}\xspace}
\mathchardef\Upsilon="7107
\def\Y#1S{\ensuremath{\Upsilon{(#1S)}}\xspace}
\def\mbc    {\mbox{$M_{\rm bc}$}\xspace}
\def\DeltaE {\mbox{$\Delta E$}\xspace}
\def\cm   {\ensuremath{{\rm \,cm}}\xspace}
\def\invfb{\ensuremath{\mbox{\,fb}^{-1}}\xspace}

\def\to{\ensuremath{\rightarrow}\xspace}
\def\CP {\ensuremath{C\!P}\xspace}
\def\ACP{{\ensuremath{\mathcal{A}_{\CP}}\xspace}}
\def\etal  {{\it et~al.}}
\def\nb    {\ensuremath{C_{N\!B}}\xspace}
\def\nbprim{\ensuremath{C'_{N\!B}}\xspace}
\def\nbmin {\ensuremath{C_{N\!B,{\rm min}}}\xspace}
\def\nbmax {\ensuremath{C_{N\!B,{\rm max}}}\xspace}
\newcommand{\stat}{\ensuremath{\mathrm{(stat)}}\xspace}
\newcommand{\syst}{\ensuremath{\mathrm{(syst)}}\xspace}
\newcommand{\gev}{\ensuremath{\mathrm{\,Ge\kern -0.1em V}}\xspace}
\newcommand{\mev}{\ensuremath{\mathrm{\,Me\kern -0.1em V}}\xspace}
\newcommand{\gevc}{\ensuremath{{\mathrm{\,Ge\kern -0.1em V\!/}c}}\xspace}
\newcommand{\mevc}{\ensuremath{{\mathrm{\,Me\kern -0.1em V\!/}c}}\xspace}
\newcommand{\gevcc}{\ensuremath{{\mathrm{\,Ge\kern -0.1em V\!/}c^2}}\xspace}
\newcommand{\mevcc}{\ensuremath{{\mathrm{\,Me\kern -0.1em V\!/}c^2}}\xspace}

\begin{document}


\preprint{\vbox{ \hbox{   }
                 \hbox{BELLE-CONF-1804}
}}

\title{ \quad\\[1.0cm] Study of charmless decays {\boldmath $\Bpm\to \KS\KS\hpm$ ($h=K,\pi$)} at Belle}


\noaffiliation
\affiliation{University of the Basque Country UPV/EHU, 48080 Bilbao}
\affiliation{Beihang University, Beijing 100191}
\affiliation{University of Bonn, 53115 Bonn}
\affiliation{Brookhaven National Laboratory, Upton, New York 11973}
\affiliation{Budker Institute of Nuclear Physics SB RAS, Novosibirsk 630090}
\affiliation{Faculty of Mathematics and Physics, Charles University, 121 16 Prague}
\affiliation{Chiba University, Chiba 263-8522}
\affiliation{Chonnam National University, Kwangju 660-701}
\affiliation{University of Cincinnati, Cincinnati, Ohio 45221}
\affiliation{Deutsches Elektronen--Synchrotron, 22607 Hamburg}
\affiliation{Duke University, Durham, North Carolina 27708}
\affiliation{University of Florida, Gainesville, Florida 32611}
\affiliation{Department of Physics, Fu Jen Catholic University, Taipei 24205}
\affiliation{Key Laboratory of Nuclear Physics and Ion-beam Application (MOE) and Institute of Modern Physics, Fudan University, Shanghai 200443}
\affiliation{Justus-Liebig-Universit\"at Gie\ss{}en, 35392 Gie\ss{}en}
\affiliation{Gifu University, Gifu 501-1193}
\affiliation{II. Physikalisches Institut, Georg-August-Universit\"at G\"ottingen, 37073 G\"ottingen}
\affiliation{SOKENDAI (The Graduate University for Advanced Studies), Hayama 240-0193}
\affiliation{Gyeongsang National University, Chinju 660-701}
\affiliation{Hanyang University, Seoul 133-791}
\affiliation{University of Hawaii, Honolulu, Hawaii 96822}
\affiliation{High Energy Accelerator Research Organization (KEK), Tsukuba 305-0801}
\affiliation{J-PARC Branch, KEK Theory Center, High Energy Accelerator Research Organization (KEK), Tsukuba 305-0801}
\affiliation{Forschungszentrum J\"{u}lich, 52425 J\"{u}lich}
\affiliation{Hiroshima Institute of Technology, Hiroshima 731-5193}
\affiliation{IKERBASQUE, Basque Foundation for Science, 48013 Bilbao}
\affiliation{University of Illinois at Urbana-Champaign, Urbana, Illinois 61801}
\affiliation{Indian Institute of Science Education and Research Mohali, SAS Nagar, 140306}
\affiliation{Indian Institute of Technology Bhubaneswar, Satya Nagar 751007}
\affiliation{Indian Institute of Technology Guwahati, Assam 781039}
\affiliation{Indian Institute of Technology Hyderabad, Telangana 502285}
\affiliation{Indian Institute of Technology Madras, Chennai 600036}
\affiliation{Indiana University, Bloomington, Indiana 47408}
\affiliation{Institute of High Energy Physics, Chinese Academy of Sciences, Beijing 100049}
\affiliation{Institute of High Energy Physics, Vienna 1050}
\affiliation{Institute for High Energy Physics, Protvino 142281}
\affiliation{Institute of Mathematical Sciences, Chennai 600113}
\affiliation{INFN - Sezione di Napoli, 80126 Napoli}
\affiliation{INFN - Sezione di Torino, 10125 Torino}
\affiliation{Advanced Science Research Center, Japan Atomic Energy Agency, Naka 319-1195}
\affiliation{J. Stefan Institute, 1000 Ljubljana}
\affiliation{Kanagawa University, Yokohama 221-8686}
\affiliation{Institut f\"ur Experimentelle Teilchenphysik, Karlsruher Institut f\"ur Technologie, 76131 Karlsruhe}
\affiliation{Kavli Institute for the Physics and Mathematics of the Universe (WPI), University of Tokyo, Kashiwa 277-8583}
\affiliation{Kennesaw State University, Kennesaw, Georgia 30144}
\affiliation{King Abdulaziz City for Science and Technology, Riyadh 11442}
\affiliation{Department of Physics, Faculty of Science, King Abdulaziz University, Jeddah 21589}
\affiliation{Kitasato University, Tokyo 108-0072}
\affiliation{Korea Institute of Science and Technology Information, Daejeon 305-806}
\affiliation{Korea University, Seoul 136-713}
\affiliation{Kyoto University, Kyoto 606-8502}
\affiliation{Kyungpook National University, Daegu 702-701}
\affiliation{LAL, Univ. Paris-Sud, CNRS/IN2P3, Universit\'{e} Paris-Saclay, Orsay}
\affiliation{\'Ecole Polytechnique F\'ed\'erale de Lausanne (EPFL), Lausanne 1015}
\affiliation{P.N. Lebedev Physical Institute of the Russian Academy of Sciences, Moscow 119991}
\affiliation{Faculty of Mathematics and Physics, University of Ljubljana, 1000 Ljubljana}
\affiliation{Ludwig Maximilians University, 80539 Munich}
\affiliation{Luther College, Decorah, Iowa 52101}
\affiliation{University of Malaya, 50603 Kuala Lumpur}
\affiliation{University of Maribor, 2000 Maribor}
\affiliation{Max-Planck-Institut f\"ur Physik, 80805 M\"unchen}
\affiliation{School of Physics, University of Melbourne, Victoria 3010}
\affiliation{University of Mississippi, University, Mississippi 38677}
\affiliation{University of Miyazaki, Miyazaki 889-2192}
\affiliation{Moscow Physical Engineering Institute, Moscow 115409}
\affiliation{Moscow Institute of Physics and Technology, Moscow Region 141700}
\affiliation{Graduate School of Science, Nagoya University, Nagoya 464-8602}
\affiliation{Kobayashi-Maskawa Institute, Nagoya University, Nagoya 464-8602}
\affiliation{Universit\`{a} di Napoli Federico II, 80055 Napoli}
\affiliation{Nara University of Education, Nara 630-8528}
\affiliation{Nara Women's University, Nara 630-8506}
\affiliation{National Central University, Chung-li 32054}
\affiliation{National United University, Miao Li 36003}
\affiliation{Department of Physics, National Taiwan University, Taipei 10617}
\affiliation{H. Niewodniczanski Institute of Nuclear Physics, Krakow 31-342}
\affiliation{Nippon Dental University, Niigata 951-8580}
\affiliation{Niigata University, Niigata 950-2181}
\affiliation{University of Nova Gorica, 5000 Nova Gorica}
\affiliation{Novosibirsk State University, Novosibirsk 630090}
\affiliation{Osaka City University, Osaka 558-8585}
\affiliation{Osaka University, Osaka 565-0871}
\affiliation{Pacific Northwest National Laboratory, Richland, Washington 99352}
\affiliation{Panjab University, Chandigarh 160014}
\affiliation{Peking University, Beijing 100871}
\affiliation{University of Pittsburgh, Pittsburgh, Pennsylvania 15260}
\affiliation{Punjab Agricultural University, Ludhiana 141004}
\affiliation{Research Center for Electron Photon Science, Tohoku University, Sendai 980-8578}
\affiliation{Research Center for Nuclear Physics, Osaka University, Osaka 567-0047}
\affiliation{Theoretical Research Division, Nishina Center, RIKEN, Saitama 351-0198}
\affiliation{RIKEN BNL Research Center, Upton, New York 11973}
\affiliation{Saga University, Saga 840-8502}
\affiliation{University of Science and Technology of China, Hefei 230026}
\affiliation{Seoul National University, Seoul 151-742}
\affiliation{Shinshu University, Nagano 390-8621}
\affiliation{Showa Pharmaceutical University, Tokyo 194-8543}
\affiliation{Soongsil University, Seoul 156-743}
\affiliation{University of South Carolina, Columbia, South Carolina 29208}
\affiliation{Stefan Meyer Institute for Subatomic Physics, Vienna 1090}
\affiliation{Sungkyunkwan University, Suwon 440-746}
\affiliation{School of Physics, University of Sydney, New South Wales 2006}
\affiliation{Department of Physics, Faculty of Science, University of Tabuk, Tabuk 71451}
\affiliation{Tata Institute of Fundamental Research, Mumbai 400005}
\affiliation{Excellence Cluster Universe, Technische Universit\"at M\"unchen, 85748 Garching}
\affiliation{Department of Physics, Technische Universit\"at M\"unchen, 85748 Garching}
\affiliation{Toho University, Funabashi 274-8510}
\affiliation{Tohoku Gakuin University, Tagajo 985-8537}
\affiliation{Department of Physics, Tohoku University, Sendai 980-8578}
\affiliation{Earthquake Research Institute, University of Tokyo, Tokyo 113-0032}
\affiliation{Department of Physics, University of Tokyo, Tokyo 113-0033}
\affiliation{Tokyo Institute of Technology, Tokyo 152-8550}
\affiliation{Tokyo Metropolitan University, Tokyo 192-0397}
\affiliation{Tokyo University of Agriculture and Technology, Tokyo 184-8588}
\affiliation{Utkal University, Bhubaneswar 751004}
\affiliation{Virginia Polytechnic Institute and State University, Blacksburg, Virginia 24061}
\affiliation{Wayne State University, Detroit, Michigan 48202}
\affiliation{Yamagata University, Yamagata 990-8560}
\affiliation{Yonsei University, Seoul 120-749}
  \author{A.~Abdesselam}\affiliation{Department of Physics, Faculty of Science, University of Tabuk, Tabuk 71451} 
  \author{I.~Adachi}\affiliation{High Energy Accelerator Research Organization (KEK), Tsukuba 305-0801}\affiliation{SOKENDAI (The Graduate University for Advanced Studies), Hayama 240-0193} 
  \author{K.~Adamczyk}\affiliation{H. Niewodniczanski Institute of Nuclear Physics, Krakow 31-342} 
  \author{J.~K.~Ahn}\affiliation{Korea University, Seoul 136-713} 
  \author{H.~Aihara}\affiliation{Department of Physics, University of Tokyo, Tokyo 113-0033} 
  \author{S.~Al~Said}\affiliation{Department of Physics, Faculty of Science, University of Tabuk, Tabuk 71451}\affiliation{Department of Physics, Faculty of Science, King Abdulaziz University, Jeddah 21589} 
  \author{K.~Arinstein}\affiliation{Budker Institute of Nuclear Physics SB RAS, Novosibirsk 630090}\affiliation{Novosibirsk State University, Novosibirsk 630090} 
  \author{Y.~Arita}\affiliation{Graduate School of Science, Nagoya University, Nagoya 464-8602} 
  \author{D.~M.~Asner}\affiliation{Brookhaven National Laboratory, Upton, New York 11973} 
  \author{H.~Atmacan}\affiliation{University of South Carolina, Columbia, South Carolina 29208} 
% ...
\collaboration{The Belle Collaboration}
\begin{abstract}
We report a search for charmless hadronic decays of charged $B$ mesons to the
final states $\KS\KS\kpm$ and $\KS\KS\pipm$. The results are based on a 
$711\invfb$ data sample that contains $772\times 10^6$ $\BB$ pairs, and was 
collected at the $\Y4S$ resonance with the Belle detector at the KEKB 
asymmetric-energy $e^+e^-$ collider. For $B^{\pm} \to \KS\KS\kpm$ decays, the measured branching 
fraction and direct $\CP$ asymmetry are $[10.64\pm0.49\stat\pm 0.44\syst]\times10^{-6}$ 
and [$-0.6\pm3.9\stat\pm 3.4\syst$]\%, respectively. In the absence of a statistically significant signal for $B^{\pm}\to \KS\KS\pipm$, we set the 90\% confidence-level upper limit on its branching fraction at $1.14 \times10^{-6}$.
\end{abstract}


\maketitle

\tighten
{\renewcommand{\thefootnote}{\fnsymbol{footnote}}}
\setcounter{footnote}{0}

\begin{figure}[!htb]
\begin{center}$
\begin{array}{cc}
\includegraphics[width=.4\textwidth]{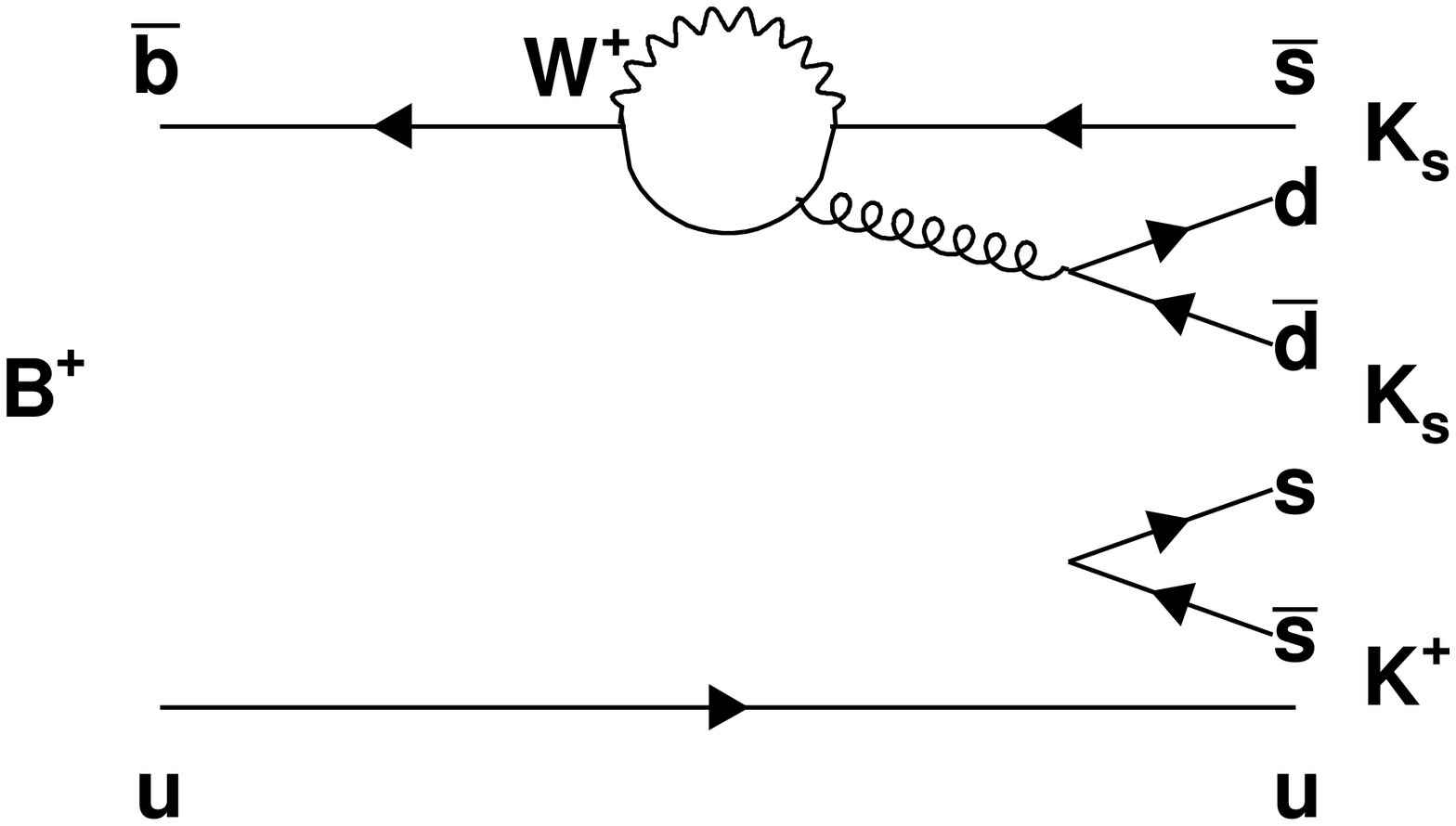} &
\includegraphics[width=.4\textwidth]{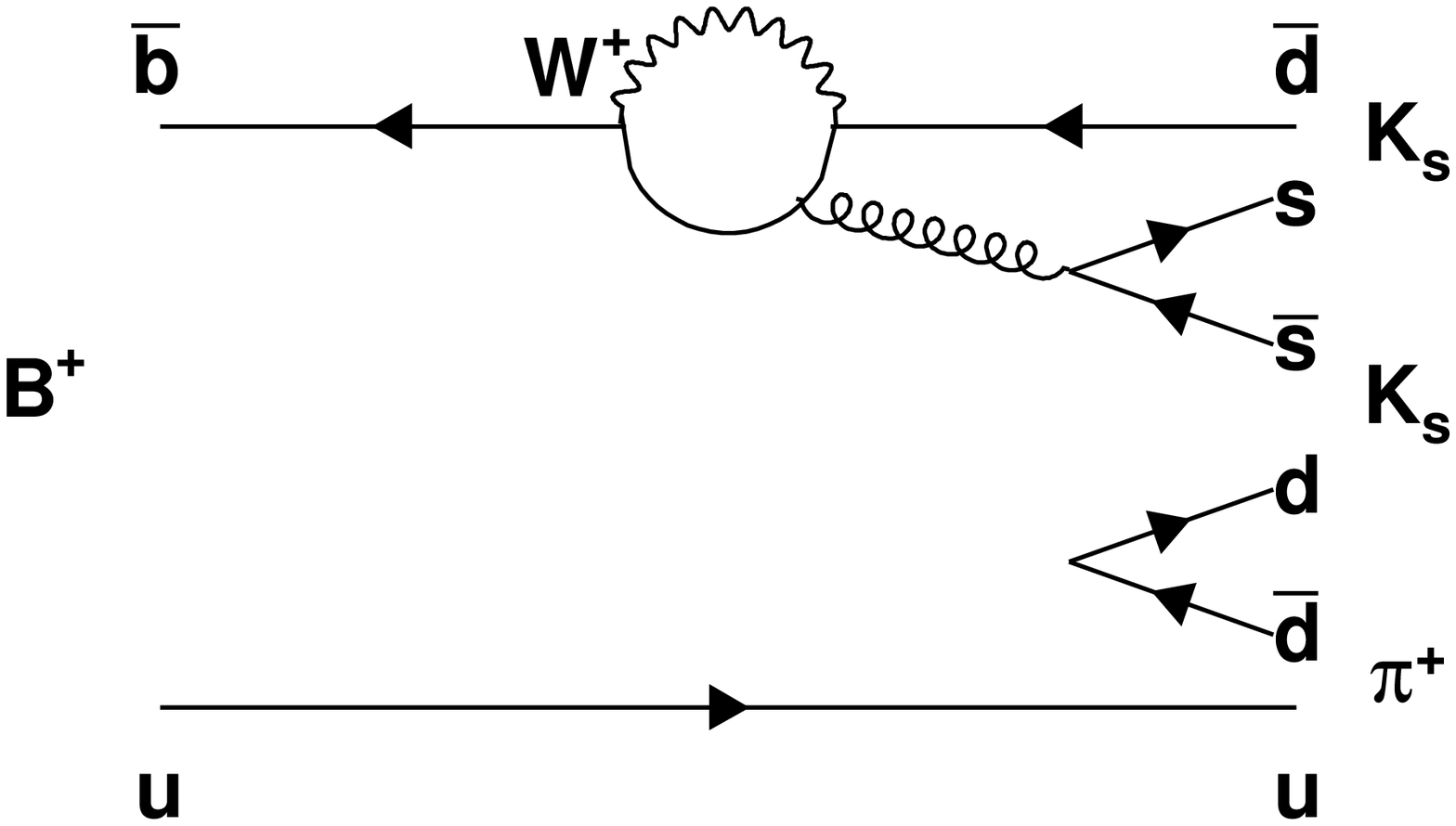} \\
\end{array}$
\end{center}
\caption{Dominant Feynman diagrams that contribute to the
 decays $B^{\pm}\to\KS\KS\kpm$ (left) and $B^{\pm}\to\KS\KS\pipm$ (right).}
\label{fig:Fey}
\end{figure}

Charged $B$-meson decays to three-body charmless hadronic 
final states $\KS\KS\kpm$ and $\KS\KS\pipm$ mainly proceed 
via the $\bbar\to\sbar$ and $\bbar\to\dbar$ loop transitions, respectively. Figure~\ref{fig:Fey} shows the
dominant Feynman diagrams that contribute to the decays. 
These are flavour changing neutral current transitions, 
which are suppressed in the standard model (SM) and hence 
provide a good avenue to search for physics beyond the 
SM~\cite{PAP:ref}. Further motivation, especially to study the contributions of various quasi-two-body 
resonances to  inclusive $\CP$ asymmetry, comes from the recent results on  $B^{\pm}\to\Kp\Km K^{\pm}$, $\Kp\Km \pi^{\pm}$ and other such three-body decays~\cite{LHCb:paper1,LHCb:paper2,Chialing:paper}. LHCb has found large inclusive asymmetries in $B^{\pm}\to\Kp\Km \pi^{\pm}$ and $\pi^+\pi^-\pi^{\pm}$ decays~\cite{LHCb:paper2}, where the observed phenomena are largely in localized regions of phase space. Recently, Belle has also reported strong evidence for a large $\CP$ asymmetry in the low $K^{+}K^{-}$ invariant-mass region of $B^{\pm}\to\Kp\Km \pi^{\pm}$~\cite{Chialing:paper}. 

The three-body decay $\Bp\to\KS\KS\Kp$~\cite{charge:ref1} has already been observed 
and subsequently studied  by the Belle and BaBar Collaborations~\cite{Belle:paper1,BaBar:paper1,BaBar:paper2}. 
Belle measured its branching fraction as 
$(13.4\pm1.9\pm1.5)\times10^{-6}$ based on a small data set of $70\invfb$~\cite{Belle:paper1}, while BaBar reported a branching fraction of $(10.6\pm0.5\pm0.3)\times10^{-6}$ and an inclusive $\CP$ 
asymmetry of $(4_{-5}^{+4}\pm2)\%$ using $426\invfb$ of data~\cite{BaBar:paper1}. The quoted uncertainties are statistical and systematic, respectively. On the other hand, the decay $\Bp\to\KS\KS\pip$ has not yet been observed, with the most restrictive upper limit being available at 90\,\% confidence level, 
${\cal B}(\Bp\to\KS\KS\pip)<5.1\times 10^{-7}$, from BaBar~\cite{BaBar:paper2}.

We present herein an improved measurement of the branching fraction and direct $\CP$ asymmetry  of the decay $\Bp\to\KS\KS \Kp$ as well as a search for the decay $\Bp\to\KS\KS\pip$  based on the full $\Upsilon(4S)$ data sample, containing $772\times 10^6$  $\BB$ pairs, collected with the Belle
detector~\cite{Belle} at the KEKB asymmetric-energy $e^+e^-$ ($3.5$
on $8.0\gev$) collider~\cite{KEKB}. The direct $\CP$ asymmetry in the former case is given by 
\begin{eqnarray}
 \ACP = \dfrac{N (\Bm\to\KS\KS \Km) - N(\Bp\to\KS\KS \Kp)}{N(\Bm\to\KS\KS \Km) + N (\Bp\to\KS\KS \Kp)}, 
\end{eqnarray}
where $N$ is the signal yield obtained for the corresponding mode. The principal detector components used in the study are: a silicon vertex detector, a $50$-layer central drift chamber (CDC), an array of aerogel threshold Cherenkov counters (ACC), a barrel-like arrangement of time-of-flight scintillation counters (TOF), and a CsI(Tl) crystal electromagnetic calorimeter (ECL). All these components are located inside a $1.5$\,T solenoidal
magnetic field.

To reconstruct $\Bp\to\KS\KS\hp$ decay candidates, we combine a pair 
of $\KS$  mesons with a charged kaon or pion. Each charged track candidate must 
have a distance of closest approach with respect to the interaction point (IP) 
of less than $0.2\cm$ in the transverse $r$--$\phi$ plane and less than 
$5.0\cm$ along the $z$ axis. Here, the $z$ axis is the direction
opposite the $e^+$ beam. Charged kaons and pions are identified based on a 
likelihood ratio ${\cal R}_{K/\pi}={{\cal L}_K}/({\cal L}_K+{\cal L}_\pi)$, 
where ${\cal L}_K$ and ${\cal L}_\pi$ denote the individual likelihood 
for kaons and pions, respectively, calculated using specific ionization 
in the CDC and information from the ACC and the TOF. A requirement, 
${\cal R}_{K/\pi}>0.6$, is applied to select the kaon candidates; track candidates 
failing it are classified as pions. The efficiency for kaon 
(pion) identification is $86\%$ ($91\%$) with a pion (kaon) misidentification rate of 
about $14\%$ ($9\%$).

The $\KS$ candidates are reconstructed from pairs of oppositely charged 
tracks, both treated as pions, and are identified with a neural network 
(NN)~\cite{neurobayes}. The NN uses the following seven input variables: the 
$\KS$ momentum in the laboratory frame, the distance along the $z$ axis 
between the two track helices at their closest approach, the $\KS$ flight length 
in the transverse plane, the angle between the $\KS$ momentum and the vector 
joining the IP to the $\KS$ decay vertex, the angle between the pion 
momentum and the laboratory frame direction in the $\KS$ rest frame, the 
distances of closest approach in the transverse plane between the IP and the two 
pion helices, and the total number of hits (in the CDC and SVD) for each 
pion track. We also require that the reconstructed invariant mass be 
between $491$ and $505 \mevcc$, corresponding to $\pm 3\sigma$ 
around the nominal $\KS$ mass~\cite{PDG}. 

$B$ meson candidates are identified using two kinematic variables:
beam-energy constrained mass, $\mbc=\sqrt{E^2_{\rm beam}/c^{4}-\left|\sum_{i}
\vec{p}_i/c\right|^2}$, and energy difference, $\DeltaE=\sum_{i}E_{i}-
E_{\rm beam}$, where $E_{\rm beam}$ is the beam energy, and $\vec{p}_i$
and $E_i$ are the momentum and energy, respectively, of the $i$-th
daughter of the reconstructed $B$ candidate in the center-of-mass (CM) frame.
We retain events with $5.271 \gevcc<\mbc<5.287\gevcc$ and $-0.10\gev<\DeltaE<0.15\gev$ 
for further analysis. The $\mbc$ requirement corresponds to approximately 
$\pm 3\sigma$ around the nominal $\Bp$ mass~\cite{PDG}. We apply a 
looser ($-6\sigma$,\,$+9\sigma$) requirement on $\DeltaE$ as it 
is used in the fitter (described below). The average number of $B$ 
candidates found per event is $1.13$ ($1.49$) for $\Bp\to\KS\KS\Kp$ 
($\KS\KS\pip$). In events with multiple $B$ candidates, we choose 
the one with the lowest $\chi^2$ value obtained from a $B$ vertex fit.
This criterion selects the correct $B$-meson candidate in
75\% (63\%) of MC events for $\Bp\to\KS\KS\Kp$ ($\KS\KS\pip$).

The dominant background is from the $e^+e^-\to\qqbar$ ($q=u,d,s,c$)
continuum process. To suppress it, observables based on
the event shape topology are utilized. The event shape in the CM frame is
expected to be spherical for $\BB$ events,  in contrast to jet-like for continuum
events. We employ another NN~\cite{neurobayes} to combine the
following six input variables: the Fisher discriminant formed from
$16$ modified Fox-Wolfram moments~\cite{KSFW}, the cosine of the angle
between the $B$ momentum and the $z$ axis, the cosine of the angle between
the $B$ thrust and the $z$ axis, the cosine of the angle between the
thrust axis of the $B$ candidate and that of the rest of the event, the
ratio of the second to the zeroth order Fox-Wolfram moments, and the vertex separation along the
$z$ axis between the $B$ candidate and the remaining tracks. The first five quantities are calculated in the CM frame. The NN training and optimization are performed with signal and
$\qqbar$ Monte Carlo (MC) simulated events. The signal MC sample is
generated with the {\textsc EvtGen} program~\cite{evtgen} assuming a
three-body phase space. We require the NN output ($\nb$)
to be greater than $-0.2$ to substantially reduce the continuum
background. The relative signal efficiency due to this requirement is
approximately $91\%$, whereas the achieved continuum suppression is
close to $84\%$ for both decays. The remainder of the $\nb$ distribution strongly peaks
near $1.0$ for signal, making it difficult to model it with
an analytic function. However, its transformed variable
\begin{eqnarray}
\nbprim=\log\left[\frac{\nb-\nbmin}{\nbmax-\nb}\right], 
\end{eqnarray}
where $\nbmin=-0.2$ and $\nbmax\simeq 1.0$, has a Gaussian-like distribution.

The background due to $B$ decays mediated via the dominant $b\to c$ transition 
is studied with an MC sample comprising such decays. The  resulting $\DeltaE$ and $\mbc$ distributions are found to strongly peak in the signal region for both $\Bp\to\KS\KS \Kp$ and  $\KS\KS \pip$ decays. For $\Bp\to\KS\KS\Kp$, the peaking background predominantly stems from $\Bp\to\Dz \Kp$ with $\Dz\to \KS\KS$ and $\Bp\to\chi_{c0}(1P) \Kp$ with $\chi_{c0}(1P)\to \KS\KS$.
To suppress these backgrounds, we exclude candidates for which $\mkk$ lies in the ranges of $[1.85,1.88]\gevcc$ and $[3.38,3.45]\gevcc$ corresponding to about 
$\pm 3\sigma$ window around the nominal $D^0$ and $\chi_{c0}(1P)$ mass~\cite{PDG}, respectively. On the other hand, in case of $\Bp\to\KS\KS\pip$, the peaking background largely arises from $\Bp\to\Dz \pip$ with $\Dz\to \KS\KS$. To suppress this background, we 
exclude candidates for which $\mkk$ lies in the aforementioned $D^{0}$ mass window. 

There are a few background modes that contribute in the $\mbc$ signal
region but have the $\DeltaE$ peak shifted from zero on the positive 
(negative) side for $\Bp\to\KS\KS\Kp$ ($\KS\KS\pip$). The so-called 
``feed-across background'' modes, mostly arising due to $K$--$\pi$ 
misidentification, are identified with a $\BB$ MC sample in which one of 
the $B$ mesons decays via $b\to u,d,s$ transitions. The feed-across background includes contribution from 
$B\to\KS\KS\pi$ ($\KS\KS K$) in $\Bp\to\KS\KS\Kp$ ($\KS\KS\pip$). 
The events that remain after removing the signal and feed-across components 
comprise the ``combinatorial background.'' After all selection requirements, the efficiency for correctly reconstructed signal events ($\epsilon_{\mathrm{rec}}$) is 24\% (28\%) for $\Bp\to\KS\KS\Kp$ ($\KS\KS\pip$). The fraction of misreconstructed signal events ($f_{\mathrm{SCF}}$) is 0.45\% (1.05\%) for $\Bp\to\KS\KS\Kp$ ($\KS\KS\pip$). As $f_{\mathrm{SCF}}$ represents a small fraction of the signal events for both decays, we consider it as a part of signal. Note that $\epsilon_{\mathrm{rec}}$ and $f_{\mathrm{SCF}}$ are determined with an MC simulation in which decays are generated assuming a three-body phase space.

The signal yield and $\ACP$ are obtained with an unbinned extended maximum 
likelihood fit to the two-dimensional distributions of $\DeltaE$ and $\nbprim$. 
We define a probability density function (PDF) for each event category $j$
(signal, $\qqbar$, combinatorial $\BB$, and feed-across backgrounds) as 
\begin{eqnarray}
  {\cal P}_{j}^{i}\equiv\dfrac{1}{2} (1-q^{i}.{\ensuremath{\mathcal{A}_{\CP ,j}}})\times {\cal P}_j(\DeltaE^{\,i})\times{\cal P}_j(\nb'^{\,i}),
\end{eqnarray} 
where $i$ denotes the event index, $q^{i}$ is the charge of the $B$ candidate in the event 
($\pm 1$ for $\Bpm$), ${\cal P}_{j}$ is the PDF corresponding to the 
component $j$.  Since the correlation between $\DeltaE$ and $\nbprim$ is found to be negligible, the product of two individual PDFs is a good approximation for the total PDF. We apply a tight requirement on $\mbc$ 
instead of including it in the fitter as it exhibits large correlation with $\DeltaE$ for signal and feed-across components. 
The extended likelihood function is
\begin{eqnarray}
{\cal L} = \dfrac{e^{-\sum_{j} n_{j}}}{N!} \prod_{i} \Big[\sum_{j} n_{j} \mathcal{P}_{j}^{i} \Big],
\end{eqnarray}
where $n_j$ is the yield of the event category $j$ and $N$ is the total number of events. To account for  crossfeed  between the  $B\to\KS\KS K$ and  $B\to\KS\KS \pi$  channels, they are simultaneously fitted, with the $B\to\KS\KS K$  signal yield in the correctly reconstructed sample determining the normalization of the crossfeed in the $B\to\KS\KS \pi$ fit region, and vice versa. 

\begin{table}[htb]
\centering
\caption{List of PDFs used to model the $\DeltaE$ and $\nbprim$
distributions for various event categories for $B\to\KS\KS K$. G, AG, 
and Poly1 denote Gaussian, asymmetric Gaussian, and first order polynomial, respectively.\\}
\label{tab:pdf-shape}
\begin{tabular}{lcccc}
\hline\hline
Event category & & $\DeltaE$ & & $\nbprim$ \\
\hline
Signal & & 3 G+Poly1 & & G+AG \\
Continuum $\qqbar$ & & Poly1 & & 2 G \\
Combinatorial $\BB$ & & Poly1 & & 2 G \\
Feed-across & & G+Poly1 & & G \\
\hline\hline
\end{tabular}
\end{table} 

Table~\ref{tab:pdf-shape} lists the PDF shapes used to model $\DeltaE$ and
$\nbprim$ distributions for various event categories for $B\to\KS\KS K$. 
For $B\to\KS\KS \pi$, we use similar PDF shapes except for the feed-across
background component, where we use a sum of a Gaussian, asymmetric 
Gaussian and first order polynomial to parametrize 
$\DeltaE$, and a sum of  Gaussian and asymmetric Gaussian functions to 
parametrize $\nbprim$. For $B\to\KS\KS K$, the yields for all event 
categories except for that of the combinatorial $\BB$ background are 
allowed to vary in the fit. The latter yield is fixed to the MC value as it is found to be correlated with the continuum background yield. For $B\to\KS\KS \pi$, the yields for all 
event categories are allowed to vary. For both $B\to\KS\KS K$ and 
$\KS\KS\pi$, the following PDF shape parameters of the continuum background
are floated: the slope of the first order polynomial used for $\DeltaE$, and 
one of the means and widths of the Gaussian functions used to
model $\nbprim$. The PDF shapes for signal and other background
components are fixed to the corresponding MC expectations. Shared parameters in the simultaneous
fit are the signal yields of $\KS\KS K$ and $\KS\KS\pi$. The ratio of the $\KS\KS K$ feed-across to 
the signal $\KS\KS\pi$ yield is floated, whereas the ratio of the $\KS\KS\pi$ 
feed-across to the signal $\KS\KS K$ yield is fixed in the fitter. This is because the latter contribution is small. We correct the signal $\DeltaE$ and $\nbprim$ 
PDF shapes for possible data-MC differences, according to the values 
obtained with a large-statistics control sample of $B\to\Dz(\KS\pip\pim)\pi$. 
The same correction factors are also applied for the feed-across 
background component of $B\to\KS\KS\pi$. The stability of the two-dimensional simultaneous fit is checked via ensemble tests using both PDF-sampled and simulated MC events.

\begin{figure}
\includegraphics[width=0.488\columnwidth]{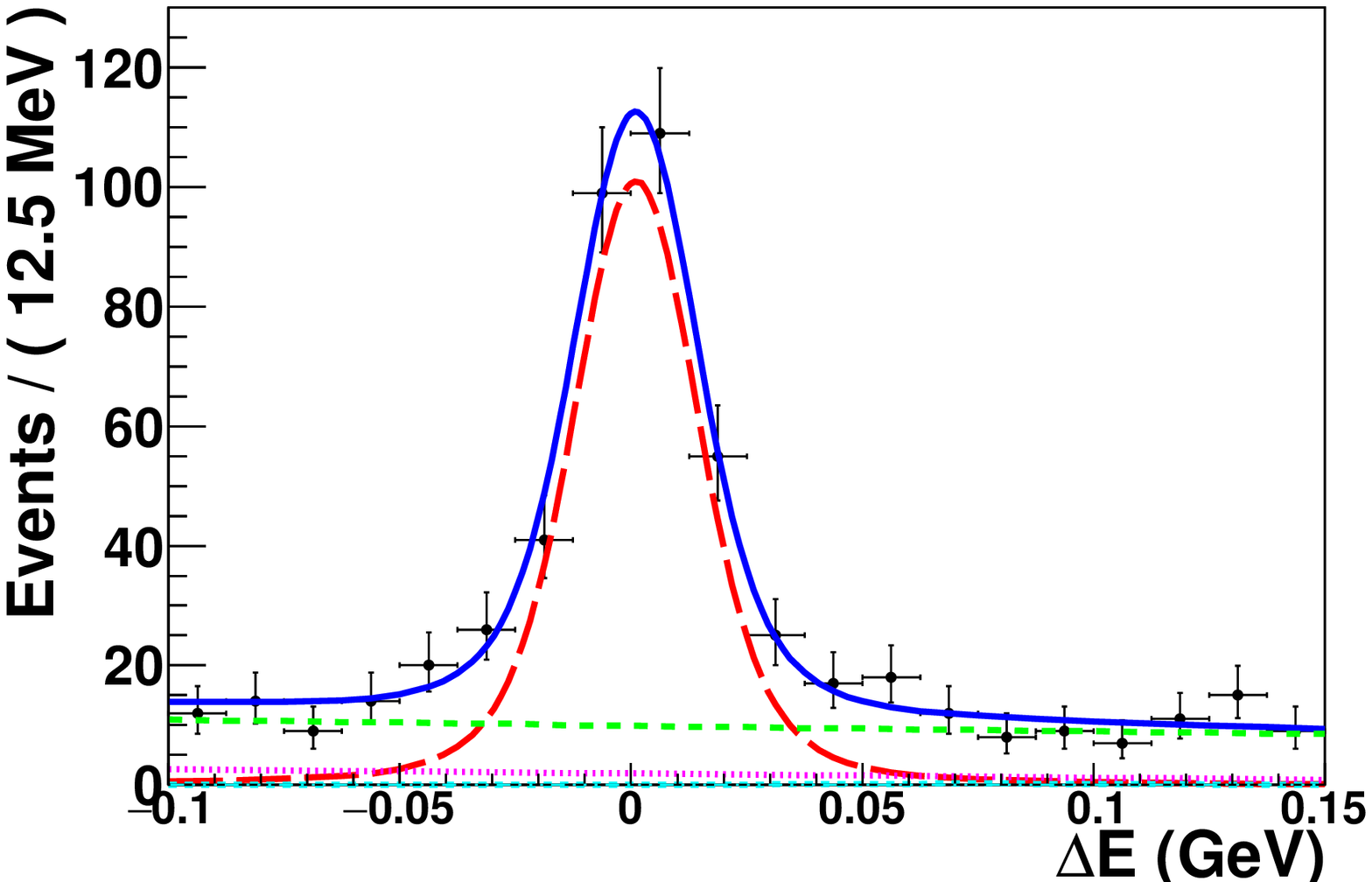}
\includegraphics[width=0.492\columnwidth]{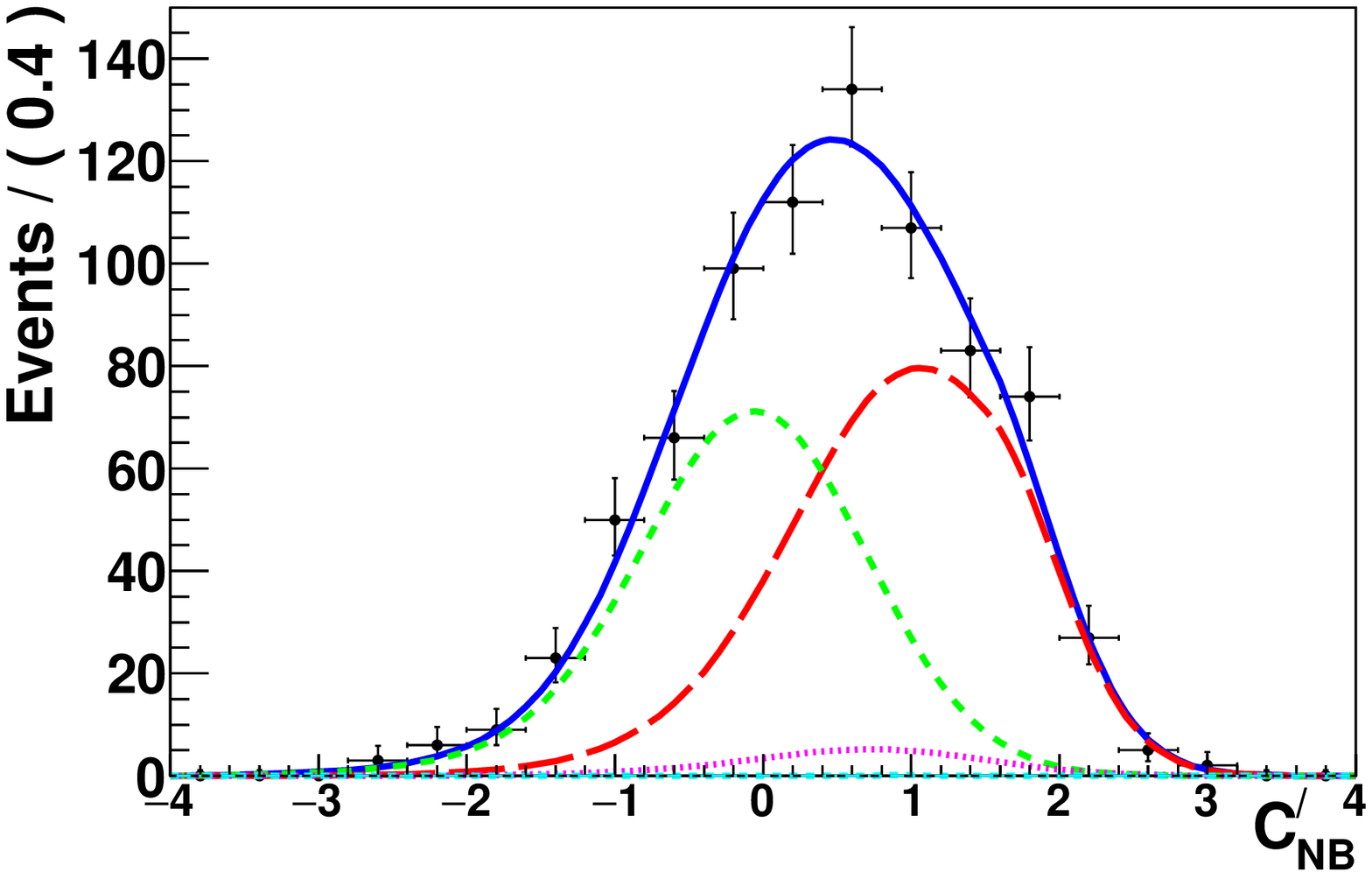}\\
a) $B^{+}\to\KS\KS K^{+}$ \\
\vspace{0.03 in}
\includegraphics[width=0.488\columnwidth]{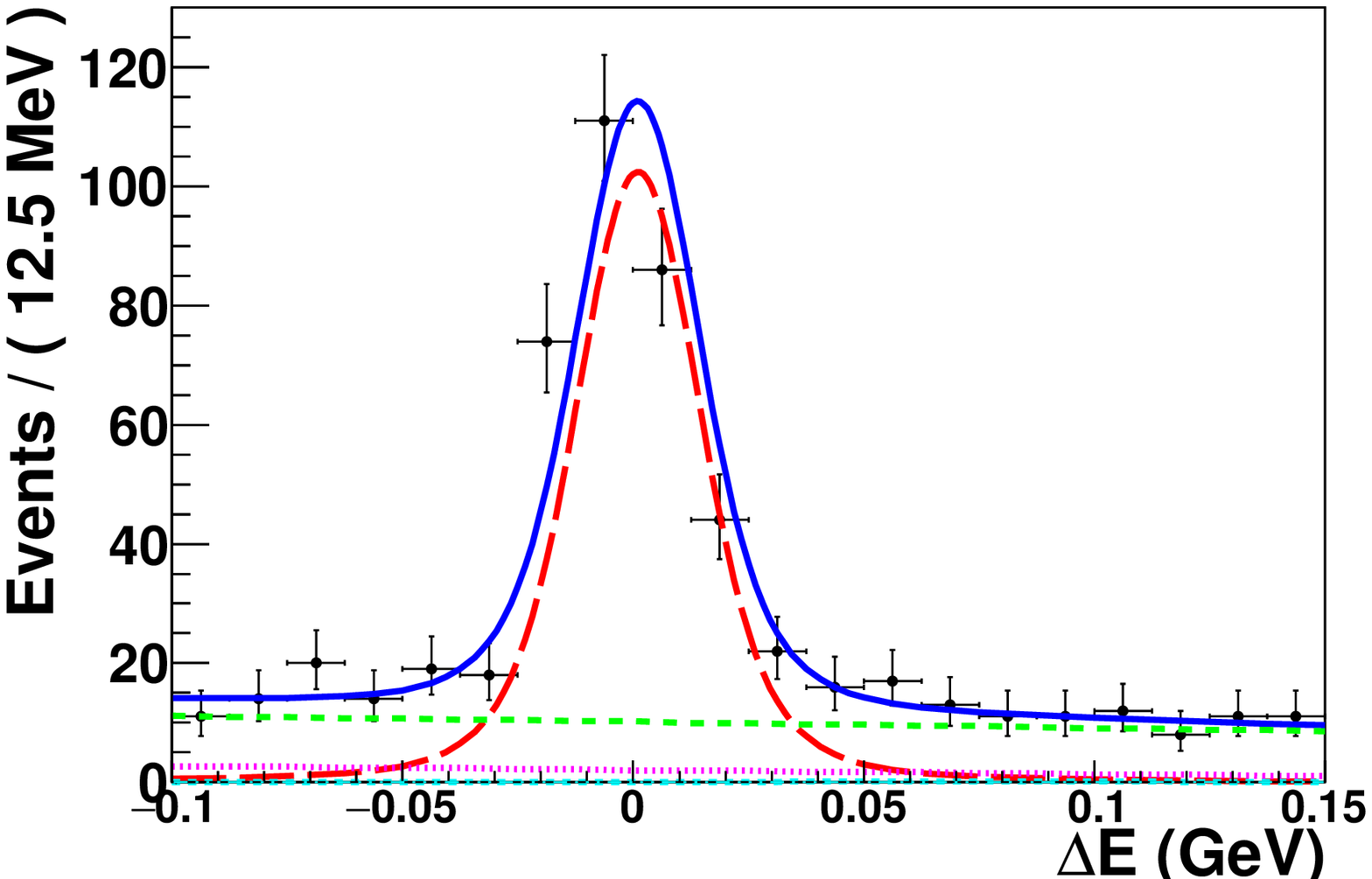}
\includegraphics[width=0.492\columnwidth]{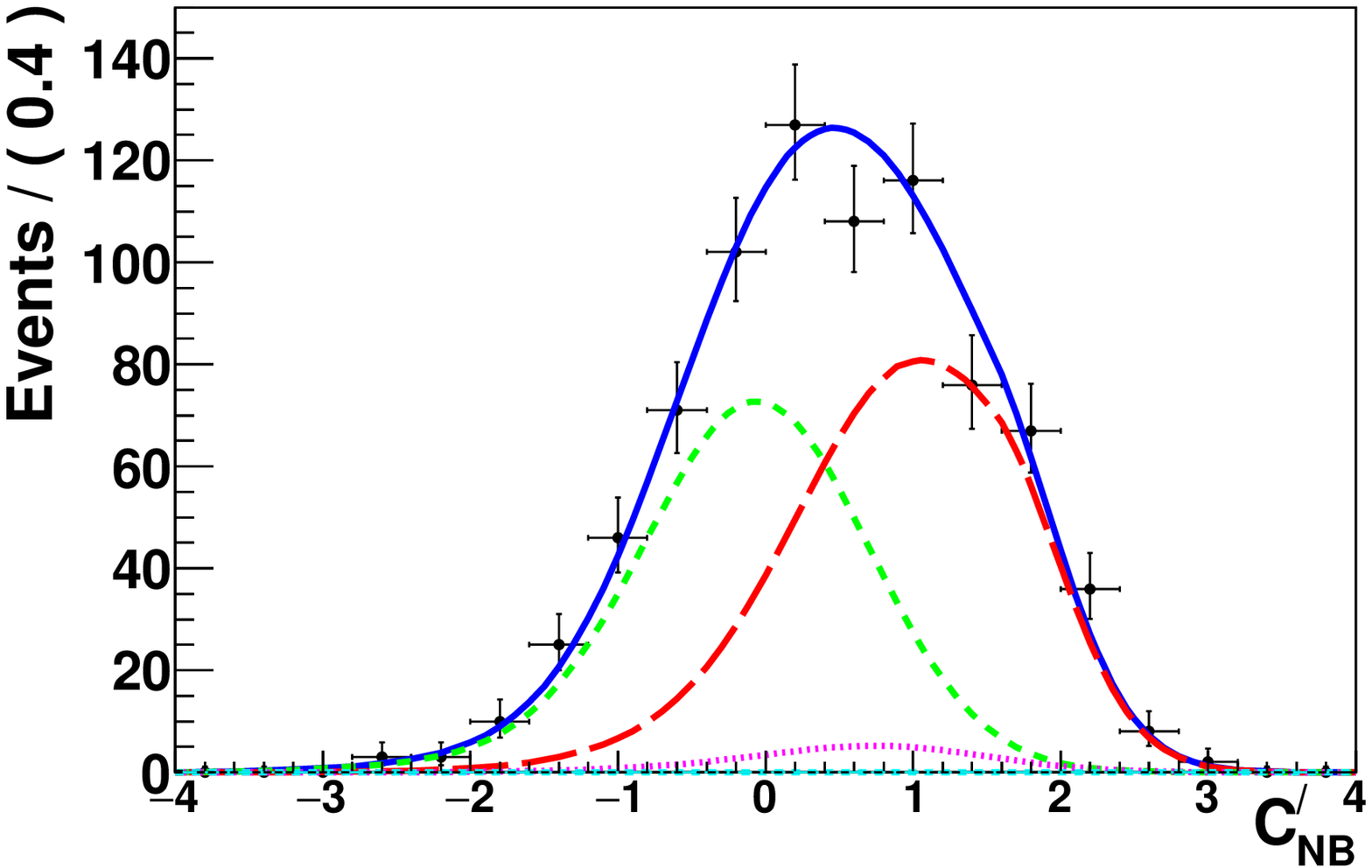}\\
 b) $B^{-}\to\KS\KS K^{-}$ \\
 \vspace{0.03 in}
\includegraphics[width=0.492\columnwidth]{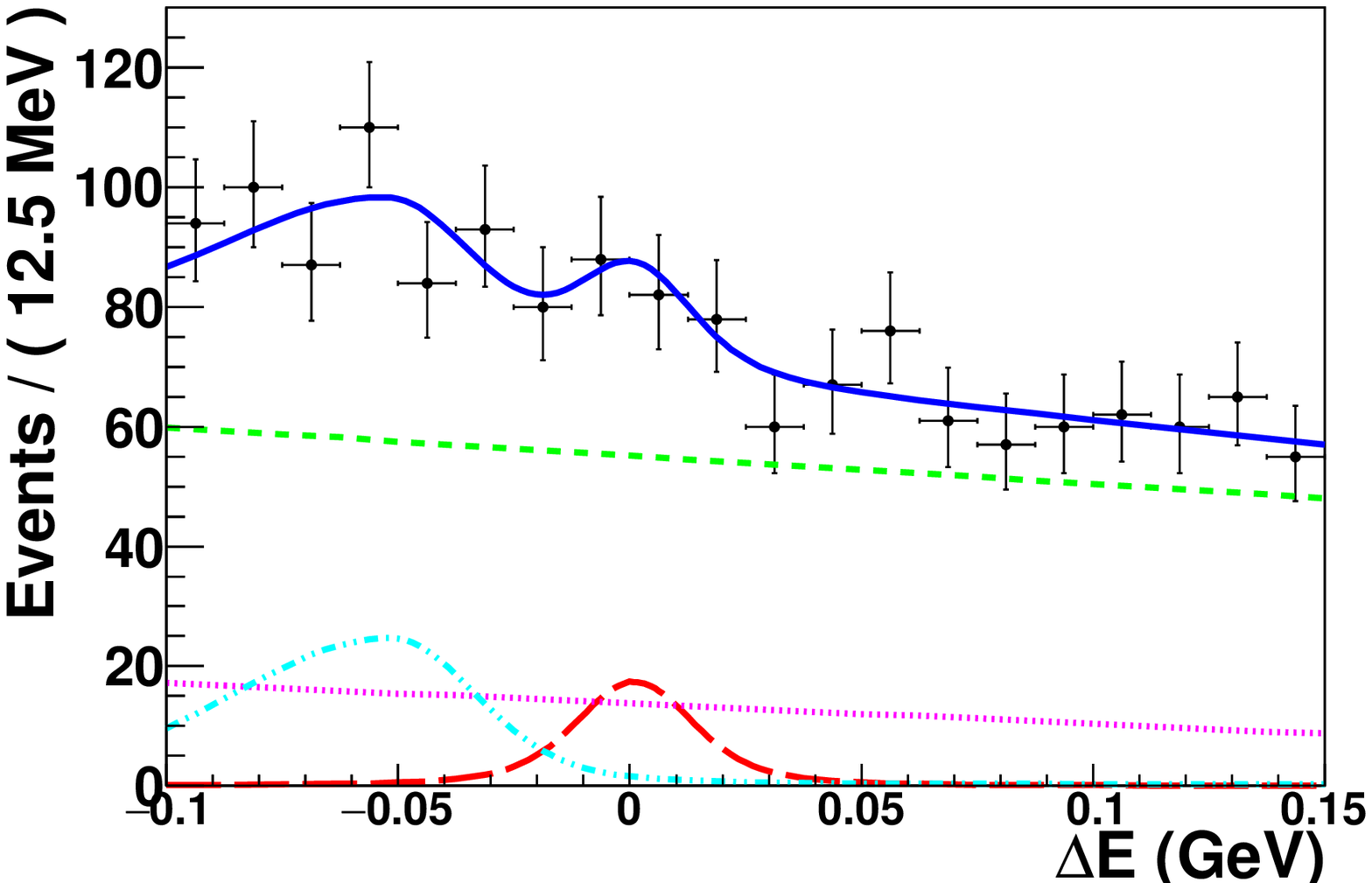}
\includegraphics[width=0.492\columnwidth]{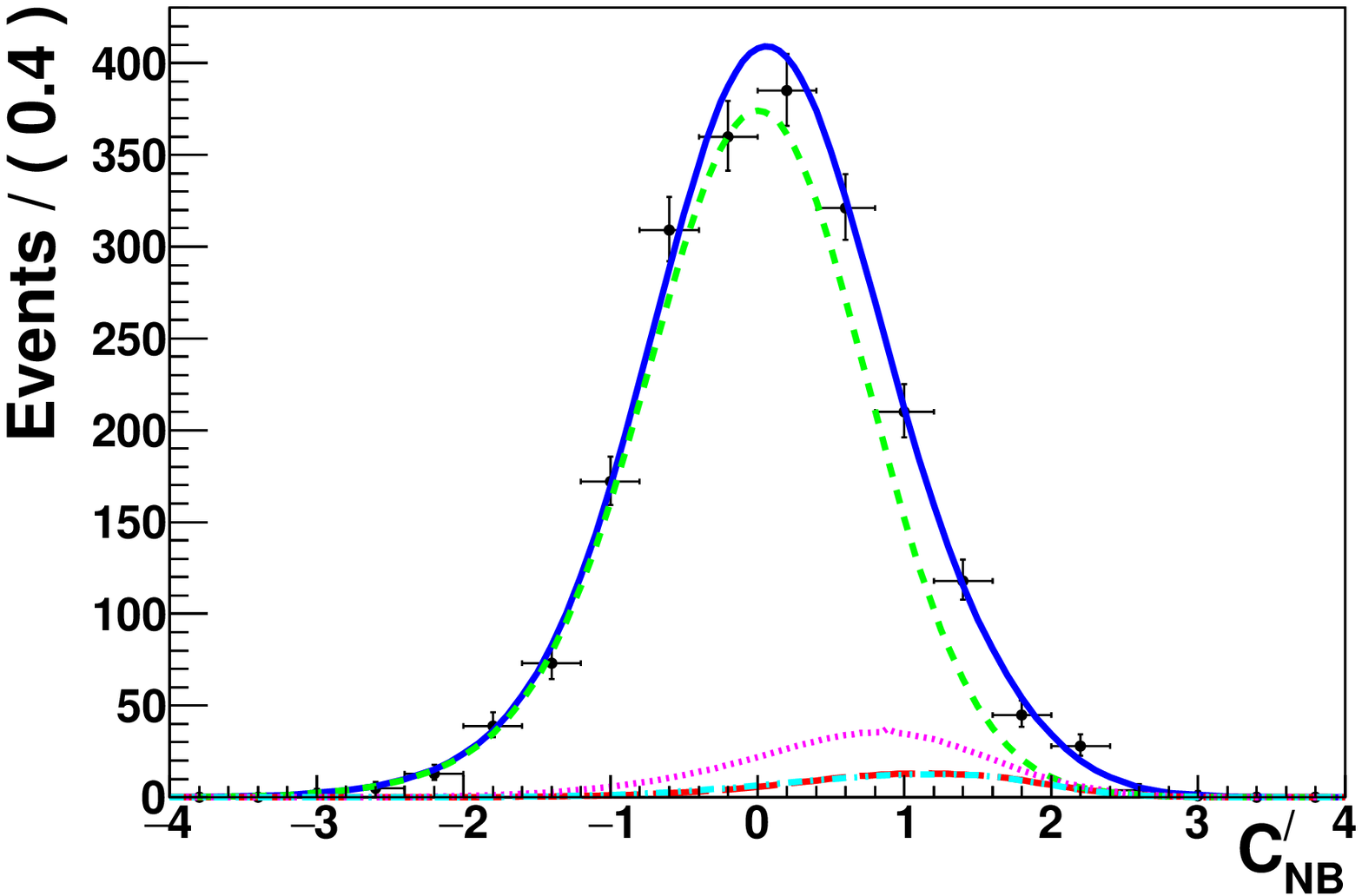}\\
\vspace{0.03 in}
 c) $B^{\pm}\to\KS\KS \pi^{\pm}$ \\
\caption{(colour online). Projections of two-dimensional simultaneous fit to $\DeltaE$ for $\nbprim>0.0$ and $\nbprim$ for $|\DeltaE|<50\mev$. Points with error bars are the data, solid blue curves are the total PDF, long dashed red curve is the signal component, dashed green curve is the continuum $\qqbar$ background, dotted magenta curve is the combinatorial $\BB$ background and dash-dotted cyan curve is the feed-across background.}
\label{fig:2D}
\end{figure}
Figure~\ref{fig:2D} shows $\DeltaE$ and $\nbprim$ projections of the fit to $B^{+}$ and $B^{-}$ samples separately for $B\to K^0_SK^0_S K$ and overall fit for $B\to K^0_SK^0_S\pi$.  We determine the branching fraction as,
\begin{equation}
 \mathcal{B}(B^{+} \rightarrow K_{S}^{0} K_{S}^{0} h^{+}) = \dfrac{N_{\mathrm{sig}}}{\epsilon  \times N_{B\bar{B}} \times [\mathcal{B}(K^{0}_{S} \rightarrow \pi \pi)]^{2}  }  
 \end{equation}
 where, $N_{\mathrm{sig}}$, $\epsilon$ and $N_{B \bar{B}}$ are the signal yield, corrected reconstruction efficiency and total number of $B \bar{B}$ pairs, respectively. For $B^{+}\to\KS\KS\pi^{+}$, we obtain a signal yield of $69\pm 26$, where the error is statistical only. The inclusive branching fraction for $B^{+}\to\KS\KS\pi^{+}$  is $(0.70 \pm 0.26 \pm 0.07)\times 10^{-6}$, where the first uncertainty is statistical and the second is systematic. Its signal significance is estimated as $\sqrt{2\log({\cal L}_0/{\cal L}_{\rm max})}$, where ${\cal L}_0$ and ${\cal L}_{\rm max}$ are the likelihood value with the signal yield set to zero and for the nominal case, respectively. Including systematic uncertainties (described below), we determine the significance to be $2.6$ standard deviations ($\sigma$). In view of the significance being less than 3$\sigma$, we set an upper limit (UL) on the branching fraction of $B\to K^0_SK^0_S\pi$. For this purpose we convolve the likelihood with a Gaussian function of width equal to the systematic error. Assuming a flat prior we set an UL of $1.14 \times 10^{-6}$ at 90\% confidence level. Our limit is somewhat looser than that of BaBar~\cite{BaBar:paper2} owing to our comparatively larger signal yield.
 
\begin{figure}[H]
\includegraphics[width=0.493\columnwidth]{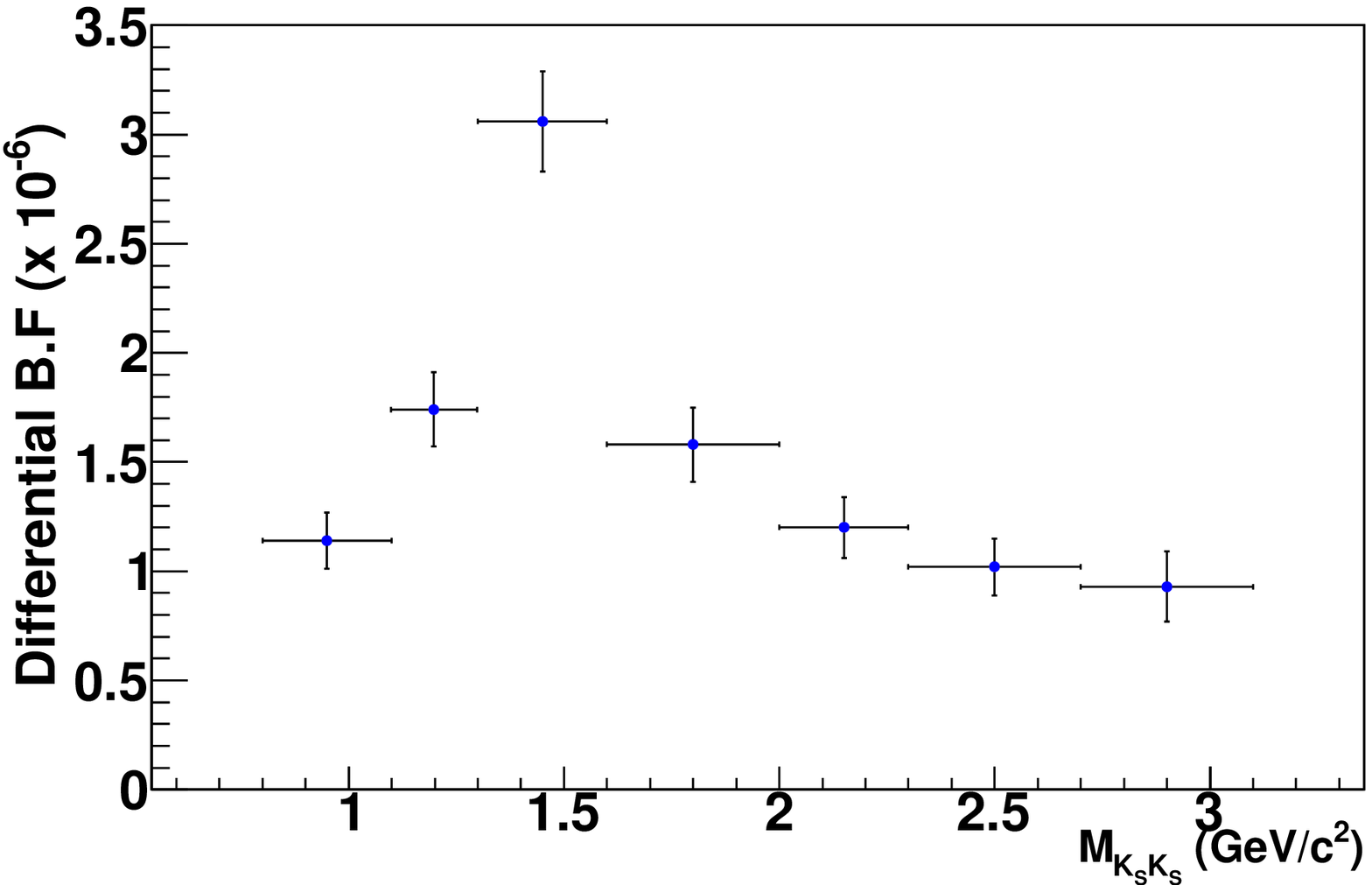} 
\includegraphics[width=0.493\columnwidth]{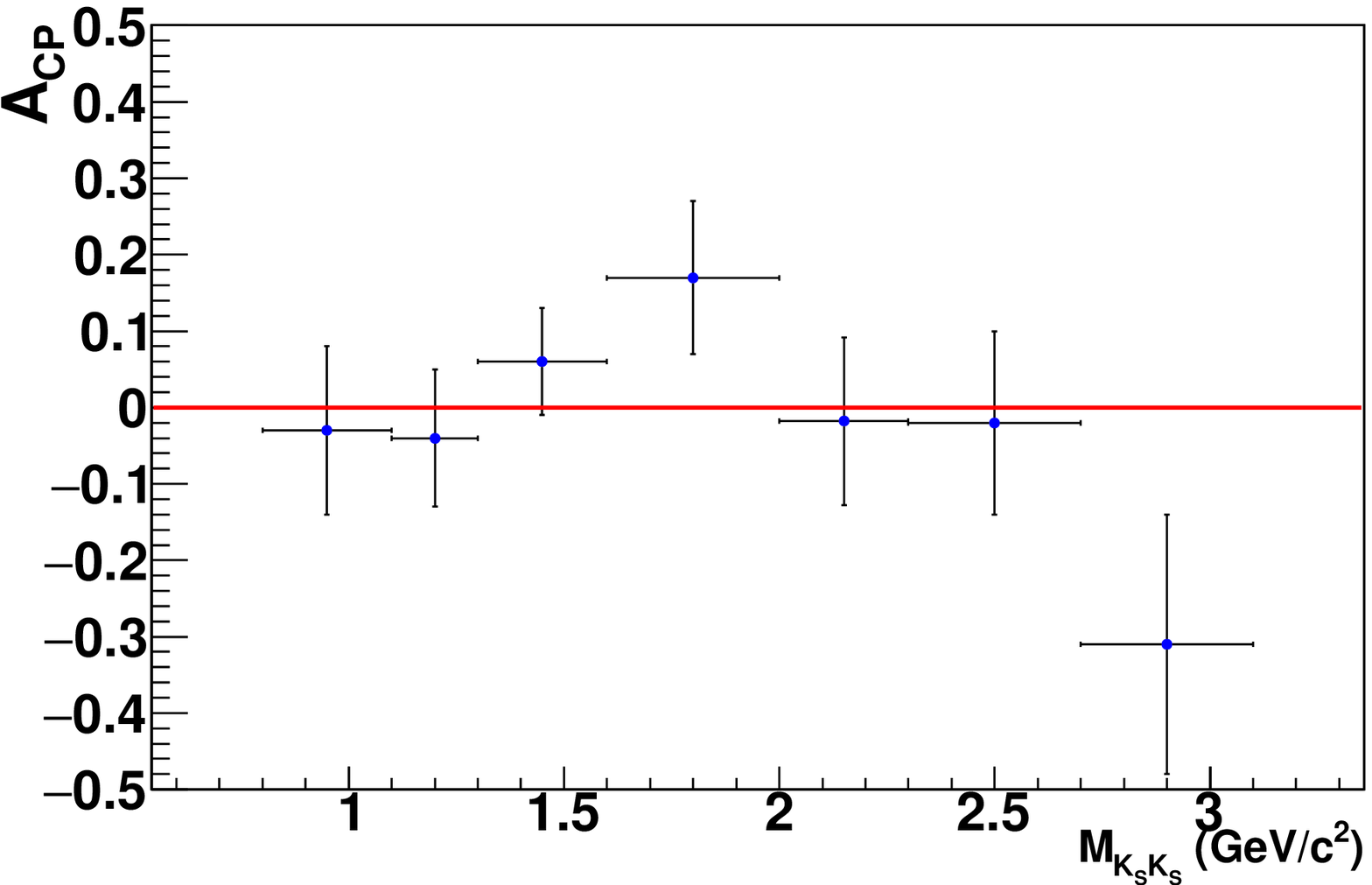} \\
\caption{
Differential branching fraction~(left) and  $\ACP$~(right) for $B^{+}\to\KS\KS K^{+}$ as a function of $\mkk$. The horizontal red line indicates null $\CP$ asymmetry.}
\label{fig:binfit}
\end{figure}

For $B^{+}\to\KS\KS K^{+}$, we perform the fit in seven bins of $\mkk$  to incorporate contributions from possible two-body intermediate resonances. Efficiency, signal yield, differential branching fraction, and $\ACP$ thus obtained  are listed in Table~\ref{tab:binfit}. Figure~\ref{fig:binfit} shows the branching fraction and $\ACP$ plotted as a function of $\mkk$. We observe an excess of events around $1.5 \gevcc$, whereas no significant evidence for $\CP$ asymmetry is found in any of the bins. The inclusive branching fraction obtained by  integrating the differential branching fractions over the entire $\mkk$ range is
\begin{equation}
\mathcal{B}(B^{+} \rightarrow K_{S}^{0} K_{S}^{0} K^{+})=(10.64 \pm 0.49 \pm 0.44)\times 10^{-6}\mbox{,}
\end{equation}
where the first uncertainty is statistical and the second  is systematic. Similarly, the inclusive $\ACP$ over the full $\mkk$ range is
\begin{equation}
\ACP(B^{} \rightarrow K_{S}^{0} K_{S}^{0} K^{}) = (-0.6 \pm 3.9 \pm 3.4)\% \mbox{.}
\end{equation}
This is obtained by weighting the $\ACP$ value in each bin with the fitted yield divided by the detection efficiency in that bin. As the statistical uncertainties are bin-independent, their total contribution is a quadratic sum. On the other hand, for the systematic uncertainties, the total contribution from the bin-correlated sources is taken as a linear sum while that from the bin-uncorrelated sources is determined as a quadratic sum. The results are in agreement with BaBar~\cite{BaBar:paper1}, where they had reported an overall $\ACP$ consistent with zero, and the presence of intermediate resonances $f_0(1500)$ and $f'_2(1525)$ in the aforementioned invariant-mass region.
 
\begin{table}[H]
\caption{Signal yield, efficiency, differential branching fraction, and $\ACP$ for each $\mkk$ bins. }
\label{tab:binfit}
\begin{center}
\begin{tabular}{c c c c c c c c}
\hline\hline
 $\mkk$ (GeV/c$^{2}$)&~ Yield ~& ~ & ~Eff. (\%)~  & ~ & ~$\Delta \mathcal{B} \times 10^{-6}$~ & ~&~ A$_{\CP}$ (\%) ~ \\   
 \hline 
 
  $<$1.1&  $98 \pm 11$  & &  $24.0 \pm 0.4$& &$1.14 \pm 0.13 \pm 0.06$& &$-3.2 \pm 11.0 \pm 3.0$ \\

  1.1-1.3 &  $145 \pm 14$ & & $23.4 \pm 0.2$ & &$1.74 \pm 0.17 \pm 0.07$ & &$-4.4 \pm 9.1 \pm 3.1$  \\
   
  1.3-1.6 &  $250 \pm 18$  & &$22.9 \pm 0.1$ & &$3.06 \pm 0.23 \pm 0.12$ & &$+6.1 \pm 6.8 \pm 3.6$  \\

  1.6-2.0 &  $122 \pm 13$  & &$21.8 \pm 0.1$ & & $1.56 \pm 0.17 \pm 0.06$& &$+16.0 \pm 10.0 \pm 4.0$  \\
 
  2.0-2.3 &  $103 \pm 12$  & &  $24.1 \pm 0.1$& &$1.20 \pm 0.14 \pm 0.05$ & &$-1.8 \pm 11.0 \pm 2.9$  \\
  
  2.3-2.7 &  $92 \pm 12$  & &$25.2 \pm 0.1$ & & $1.02 \pm 0.13 \pm 0.04$ & &$-2.0 \pm 12.0 \pm 3.2$  \\
  
  $>$ 2.7 &  $86 \pm 15$  & & $26.3 \pm 0.0$ & &$0.91 \pm 0.16 \pm 0.04$ & &$-31.2 \pm 17.0 \pm 4.2$\\
   
\hline
\end{tabular}
\end{center}
\end{table}

Major sources of systematic uncertainties on the branching fractions are same for both  $B^{+}\to\KS\KS K^{+}$ and $\KS\KS \pi^{+}$ decays. These are listed along with their contributions in Tables~\ref{tab:sys1} and ~\ref{tab:sysm}. We use partially reconstructed $\Dstarp\to\Dz(\KS\pip\pim)\pip$ decays to assign the systematic uncertainty due to charged-track reconstruction ($0.35\%$ per track). The $\Dstarp\to\Dz(\Km\pip)\pip$ control sample is used to determine the systematic uncertainty due to the $R_{K/\pi}$ requirement. The uncertainty due to the total number of $B \bar{B}$ pairs is 1.37\%. The uncertainties due to the $M_{\mathrm{bc}}$ and continuum suppression criteria are estimated using a control sample of $B^+\rightarrow \Dz(\KS\pi^-\pi^+)\pi^+$ decays. The uncertainty arising due to the $K^{0}_{S} \rightarrow \pi^{+} \pi^{-}$ reconstruction is estimated from $D^{0}\rightarrow K^{0}_{S}K^{0}_{S}$ analysis ~\cite{Dash:paper}. Potential fit bias is checked by performing an ensemble test comprising $1000$ pseudo-experiments, where the signal component is embedded from the corresponding MC samples and PDF shapes are used to generate the dataset for other event categories. The uncertainties due to signal PDF shape parameters are estimated by varying  the correction factors (discussed earlier) by $\pm 1 \sigma$ of their error. Similarly, the uncertainties due to background PDF shape parameters are calculated by varying all fixed parameters by $\pm1\sigma$. We evaluate the uncertainty due to the fixed yields of combinatorial backgrounds by varying it up and down by its statistical error. The uncertainties due to the dependence of PDF shapes on $\mkk$ are evaluated in each $\mkk$ bin and propagated to the branching fraction measurement. The total systematic uncertainty is calculated by summing all these contributions in quadrature.
\begin{table}[H]
\caption{\label{tab:sys1}Systematic uncertainties in the branching fraction for $B \to\KS\KS \pi$.}
\begin{center}
\begin{tabular}{lcc}
\hline\hline
{Source} &  Relative uncertainty in ${\cal B}$ ($\%$) \\
\hline
Tracking & $0.35$ \\
Particle identification & $0.80$ \\
Number of $\BB$ pairs & $1.37$ \\
Continuum suppression & $0.34$ \\
Requirement on $\mbc$ & $0.03$ \\
$\KS$ reconstruction efficiency & $3.25$ \\
Fit bias & $1.86$ \\
Signal PDF & $+2.50, -0.00$ \\
Combinatorial $B\bar{B}$ PDF & $+2.42, -1.78$ \\
Feed-across PDF& $+6.47, -9.56$ \\
Fixed yields&  $0.00$   \\
\hline\hline
\end{tabular}
\end{center}
\end{table}
Systematic uncertainties on $\ACP$ are listed in Table ~\ref{tab:sysm}. The systematic errors due to the signal and  background  modeling are estimated with the same procedure as done for the branching fraction. Uncertainties due to intrinsic detector bias on charged particle detection is evaluated from the $\ACP$ value obtained using a data sample of $89.4 \invfb$ recorded $60 \mev$ below the $\Upsilon(4S)$ resonance. We obtain an asymmetry of $(-2.7 \pm 2.0)\%$ for this dataset, from which we take the absolute value of the central shift (2.7\%) as the uncertainty due to detector asymmetry. The uncertainties due to the dependence of PDF shapes on $\mkk$ are evaluated in each $\mkk$ bin and propagated to the $\ACP$ value.

\begin{table}[H]
\caption{\label{tab:sysm}Systematic uncertainties for the branching fraction and $\ACP$ in the individual bins for $B \to\KS\KS K$. ``$\dagger$'' indicates the uncertainty is $\mkk$ dependent and an ellipsis indicates a value below $0.05\%$~($0.001$) in $\cal B$~($\ACP$).}
\begin{center}
\begin{tabular}{lccccccc}
\hline\hline
{Source} &  \multicolumn{5}{c}{Relative uncertainty in ${\cal B}$ ($\%$)} \\
$\mkk$(GeV/$c^2$) &  \begin{scriptsize} $<$1.1 \end{scriptsize} & \begin{scriptsize}$1.1-1.3$   \end{scriptsize}  &  \begin{scriptsize}$1.3-1.6$ \end{scriptsize}   &  \begin{scriptsize} $1.6-2.0$  \end{scriptsize}  &  \begin{scriptsize} $2.0-2.3$  \end{scriptsize} &  \begin{scriptsize} $2.3-2.7$  \end{scriptsize} &  \begin{scriptsize} $> 2.7$ \end{scriptsize} \\
\hline
Tracking & \multicolumn{5}{c}{$0.35$} \\
Particle identification & \multicolumn{5}{c}{$0.80$} \\
Number of $\BB$ pairs & \multicolumn{5}{c}{$1.37$} \\
Continuum suppression & \multicolumn{5}{c}{$0.34$} \\
Requirement on $\mbc$ & \multicolumn{5}{c}{$0.03$} \\
$\KS$ reconstruction efficiency & \multicolumn{5}{c}{$3.22$} \\
Fit bias & \multicolumn{5}{c}{$0.59$} \\
Signal and background PDF$^\dagger$ &  $\cdots$  & $^{+0.00}_{-0.69}$  & $^{+0.00}_{-0.57}$ & $^{+0.81}_{-0.81}$ & $^{+0.00}_{-0.97}$ & $^{+0.00}_{-1.09}$ & $^{+1.16}_{-1.16}$ \\
Fixed yields$^\dagger$&  $\cdots$  & $\cdots$&  $\cdots$ & $\cdots$ & $\cdots$ & $\cdots$& $\cdots$ \\
PDF dependence on $\mkk^\dagger$ & 4.08 & 0.35 & 0.60 & 0.41 & 0.95  &0.00 & 1.75 \\
\hline\hline
\multicolumn{2}{c}{} \\
\hline\hline
Source & \multicolumn{5}{c}{Absolute uncertainties in $\ACP$} \\
$\mkk$(GeV/$c^2$) & \begin{scriptsize} $<$1.1 \end{scriptsize} & \begin{scriptsize}$1.1-1.3$   \end{scriptsize}  &  \begin{scriptsize}$1.3-1.6$ \end{scriptsize}   &  \begin{scriptsize} $1.6-2.0$  \end{scriptsize}  &  \begin{scriptsize} $2.0-2.3$  \end{scriptsize} &  \begin{scriptsize} $2.3-2.7$  \end{scriptsize} &  \begin{scriptsize} $> 2.7$ \end{scriptsize} \\
\hline
Signal and background PDF$^\dagger$ &$^{-0.000}_{+0.005}$ & $^{-0.000}_{+0.005}$ & $^{-0.003}_{+0.000}$&$^{-0.031}_{+0.000}$ & $^{-0.000}_{+0.006}$  & $^{-0.000}_{+0.002}$& $^{-0.001}_{+0.006}$ \\
Fixed yields$^\dagger$  & $\cdots$ & $\cdots$& $\cdots$& $\cdots$& $\cdots$ & $\cdots$  & $\cdots$ \\
 PDF dependence on $\mkk^\dagger$ & 0.003 &0.004 & 0.009 &  0.010 & 0.002  & 0.005 & 0.015 \\
Detector bias &  \multicolumn{5}{c}{$0.027$} \\
\hline\hline
\end{tabular}
\end{center}
\end{table}
In summary, we have reported measurements of the suppressed decays $\Bp\to\KS\KS\Kp$ and $\Bp\to\KS\KS\pip$
using the full $\Y4S$ data sample collected with the Belle detector. We perform a two-dimensional simultaneous fit to extract the signal yields of both decays. We report a 90\% upper limit on the branching fraction of $1.14 \times10^{-6}$  for the decay $\Bp\to\KS\KS\pip$. We also report the branching fraction and $\ACP$ as a function of $\mkk$ for $\Bp\to\KS\KS\Kp$.  We observe an excess of events at low $\mkk$ region, likely caused by the two-body intermediate resonances reported by BaBar ~\cite{BaBar:paper1}. An amplitude analysis with more data is needed to further elucidate the nature of these resonances. The measured inclusive branching fraction and direct $\CP$ asymmetry are 
${\cal B}(\Bp\to\KS\KS\Kp) = (10.64\pm0.49\pm0.44)\times 10^{-6}$ and $\ACP = (-0.6\pm3.9\pm3.4)\%$, respectively. These supersede Belle's earlier measurements~\cite{Belle:paper1} and constitute the most precise results to date.
\ \\ 
\ \\
\ \\



\begin{thebibliography}{99}
\bibitem{charge:ref1}
Inclusion of charge-conjugate reactions are implicit unless stated otherwise.
\end{thebibliography}
\end{document}